\documentclass[twocolumn]{openjournal}

\usepackage{enumitem}
\usepackage[most,breakable]{tcolorbox}
\usepackage{graphicx}
\usepackage{amsmath}
\usepackage{xcolor}
\usepackage{hyperref}

\definecolor{boxcolor1}{RGB}{230,240,255} 

\newtcolorbox{questionbox}[1][]{
  colback=#1,
  colframe=black,
  boxrule=1pt,
  arc=5pt,
  outer arc=5pt,
  enhanced,
  breakable
}

\shorttitle{Teaching Astronomy with Large Language Models}
\shortauthors{Ting \& O'Briain}

\begin{document}

\title{Teaching Astronomy with Large Language Models \vspace{-1.5cm}}

\author{Yuan-Sen Ting\,\href{https://orcid.org/0000-0001-5082-9536}{\includegraphics[scale=0.04]{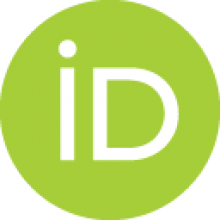}}}
\affiliation{Department of Astronomy, The Ohio State University, Columbus, OH 43210, USA}
\affiliation{Center for Cosmology and AstroParticle Physics (CCAPP), The Ohio State University, Columbus, OH 43210, USA}

\author{Teaghan O'Briain}
\affiliation{Department of Astronomy, The Ohio State University, Columbus, OH 43210, USA}

\begin{abstract}
We present a study of LLM integration in final-year undergraduate astronomy education, examining how students develop AI literacy through structured guidance and documentation requirements. We developed AstroTutor, a domain-specific astronomy tutoring system enhanced with curated arXiv content, and deployed it alongside general-purpose LLMs in the course. Students documented their AI usage through homework reflections and post-course surveys. We analyzed student evolution in AI interaction strategies and conducted experimental comparisons of LLM-assisted versus traditional grading methods. LLM grading showed strong correlation with human evaluation while providing more detailed and consistent feedback. We also piloted LLM-facilitated interview-based examinations as a scalable alternative to traditional assessments, demonstrating potential for individualized evaluation that addresses common testing limitations. Students experienced decreased rather than increased reliance on LLMs over the semester, developing critical evaluation skills and strategic tool selection. They evolved from basic assistance-seeking to verification workflows, with documentation requirements fostering metacognitive awareness. Students developed effective prompting strategies, contextual enrichment techniques, and cross-verification practices. Our findings suggest that structured LLM integration with transparency requirements and domain-specific tools can enhance astronomy education while building essential AI literacy skills. We provide implementation guidelines for educators and make our AstroTutor repository freely available.
\end{abstract}

\section{Introduction}
    
Large Language Models are transforming educational landscapes across disciplines, with astronomy education facing both opportunities and challenges. While LLMs demonstrate capabilities in explaining concepts, generating code, and assisting with mathematical derivations \citep{brown2020language, achiam2023gpt}, concerns persist about their potential to undermine critical thinking skills and create dependencies among students \citep{Kasneci2023,Zhai2024,Gerlich2025}.
    
These concerns are particularly relevant in advanced astronomy courses that are statistics-driven and coding-intensive, where students must develop problem-solving abilities, master computational techniques, and learn to critically evaluate scientific information \citep{Fouesneau2024}. The improvement of LLMs \citep{Wei2022} has posed significant challenges to classical testing approaches for most textbook coding problems, especially at the undergraduate level \citep{Kumar2023,Balalle2025}.

This transformation raises questions about which skills remain valuable to teach and assess, and how educational frameworks should adapt to prepare students for an AI-integrated workforce while maintaining core disciplinary expertise \citep{YukChan2023}. Traditional approaches to AI in education have often been binary—either complete prohibition or unrestricted access. Neither approach adequately prepares students for a future where AI tools are ubiquitous in scientific research and professional practice \citep{Boiko2023, Bran2023, Ramos2024}.
    
Recent studies in STEM education suggest that the key lies not in whether to integrate AI tools, but in how to implement them strategically to enhance rather than replace human learning processes \citep{Cooper2024,Lehmann2024,Pathak2025}. Several fields have begun exploring structured AI integration \citep{Wu2023, Chen2024, Raihan2024, Wang2024}, with implementations of AI-powered tutoring systems showing that students can access personalized instruction while maintaining learning objectives. Studies have demonstrated that AI-generated personalized hints can reduce student effort in deriving correct solutions \citep{Becker2022,Deng2025}, while meta-learning approaches have achieved feedback precision comparable to human teaching assistants \citep{Kuchemann2023}. 

In physics education specifically, controlled studies have shown that custom AI chatbots can enhance student emotions, situational interest, and self-efficacy while reducing cognitive load, though effects on learning performance remain inconclusive \citep{Lademann2025}. RAG-based Socratic tutoring approaches similar to our AstroTutor have also been explored in physics using platforms such as NotebookLM \citep{Tufino2025}, highlighting both the promise and the pedagogical tensions inherent in guided AI tutoring. However, limited documentation exists for LLM integration specifically in astronomy education, representing a gap in understanding best practices for STEM disciplines that combine theoretical knowledge with computational skills.
    
Astronomy education presents unique challenges and opportunities in this context. The field requires a blend of theoretical understanding, mathematical proficiency, and computational skills \citep{Borne2009,Feigelson2021}. Students must learn to work with large datasets, implement complex algorithms, and interpret results within the broader context of astrophysical phenomena. These requirements create a testing ground for structured LLM integration, as the discipline naturally demands the critical thinking and verification skills that effective AI usage requires.
    
We report our findings from integrating LLMs within an advanced astronomical statistics course with 12 students, most preparing for graduate school in astronomy. The course's difficulty level creates naturally self-selecting enrollment, providing a manageable teaching environment while representing a specialized sample of motivated, research-oriented students.
    
The course structure emphasized problem-solving through four homework assignments and two group projects. The homework assignments applied classical machine learning tools and statistical techniques to simplified versions of real astronomical research problems, with each assignment breaking down complex research questions into manageable, sequential steps that developed critical thinking skills. The two project-based assignments required students to engage in authentic research-like activities, with homework assignments serving as scaffolding for these independent investigations. The analysis presented in this paper focuses on the four individual homework assignments; the group projects were excluded because their collaborative nature made it difficult to attribute individual AI usage patterns to specific students.

\section{AstroTutor}

We developed AstroTutor as a domain-specific tutoring chatbot tailored for astronomy problem-solving. The system was built in December 2024 using Gemini-2.0-Flash as the backbone model\footnote{\url{https://ai.google.dev/gemini-api/docs/models}}, offering astronomy-specific knowledge at reasonable cost (approximately \$1 per student per month). We upgraded to Gemini-2.5-Flash when it became available near the course's end, though most students had already formed opinions about the tool by then. While stronger models like Gemini-2.5-Pro or Claude-3.7-Sonnet might have provided better responses, the cost difference made Gemini-Flash the practical choice for sustained educational deployment. The source code for AstroTutor is freely available on GitHub\footnote{\url{https://github.com/tingyuansen/computeastro-tutor}}.

This RAG-based agent  \citep{lewis2020} was constructed using the lecturer's own lecture notes with a ``Socratic'' teaching style, designed to guide students toward solutions through questioning and hints rather than providing direct answers. The pedagogical approach emphasized asking probing questions to help students identify knowledge gaps and develop deeper understanding through guided discovery.

\subsection{Pedagogical Design and Response Framework}

The response style was designed to mirror effective human tutoring practices through several design choices. Rather than providing direct answers, AstroTutor breaks down complex problems into smaller components, asks clarifying questions, and guides students through problem-solving processes step by step. All responses consistently end with questions to encourage continued engagement and self-reflection \citep{Wallin2018}. The system was prompted to avoid being ``overly affirming'' like many commercial LLMs, instead providing constructive feedback and gentle correction when students make errors.

AstroTutor implements active content filtering that prohibits questions outside astrostatistics and machine learning, maintaining focus on course topics and preventing distracting off-topic interactions.

Code assistance follows a tutorial-style presentation approach, alternating between code blocks and explanations rather than providing complete scripts. The system provides code in digestible chunks with accompanying explanations, focusing on concepts and approaches rather than generating entire solutions. This approach emphasizes teaching the ``why'' behind methods, not just the ``how,'' connecting topics to broader themes in astronomy and machine learning.

\subsection{Multi-Agent Architecture}

The system employs a multi-agent architecture comprising several specialized components working in coordination to provide educational support:

\begin{figure*}[ht]
    \centering
    \includegraphics[width=0.9\textwidth]{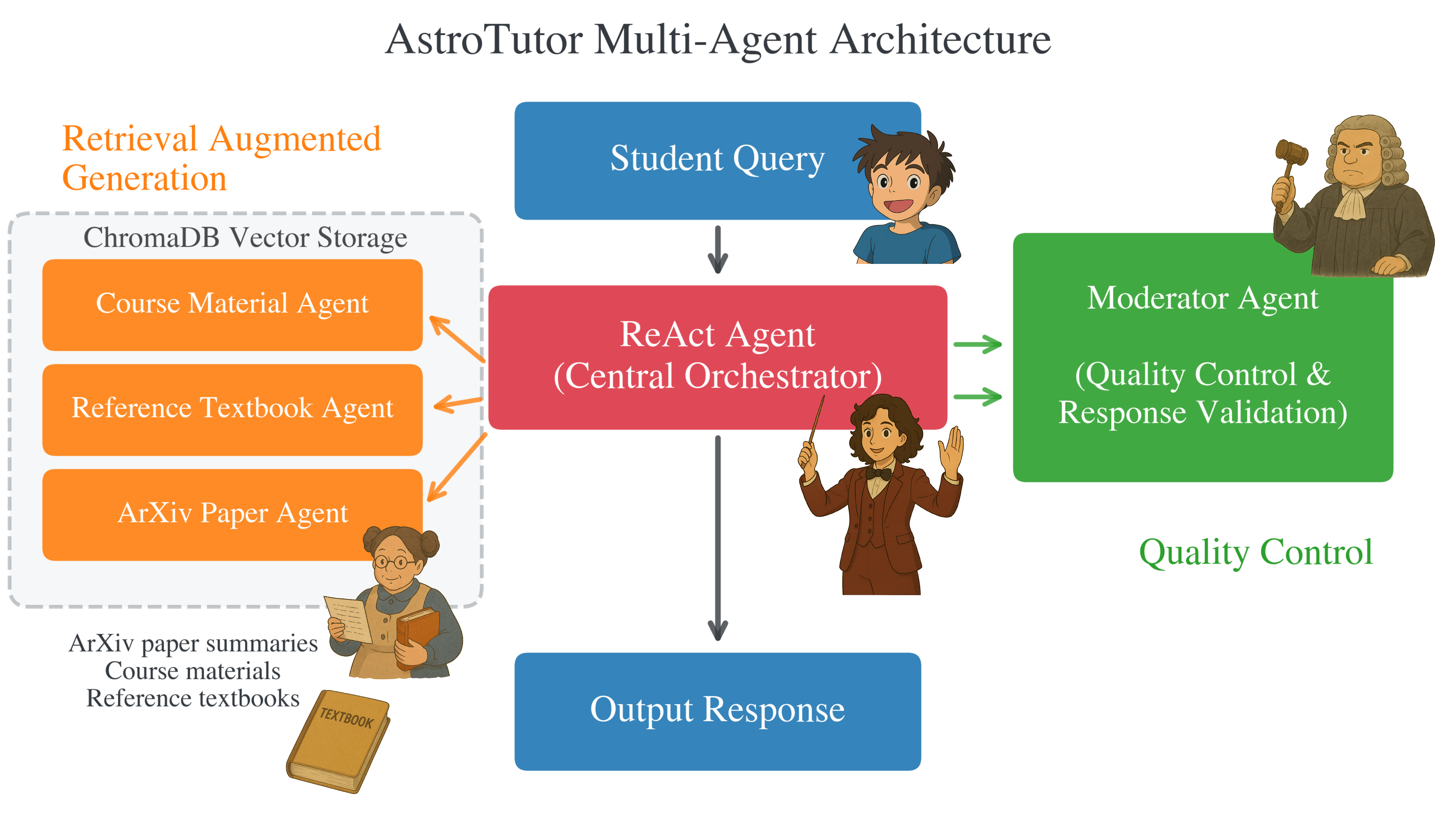}
    \caption{Multi-agent architecture of AstroTutor. The system employs a Retrieval Augmented Generation (RAG) approach with three specialized agents accessing distinct knowledge domains: course materials and lecture notes adapted from the instructor's textbook, trusted reference materials, and a curated database of ArXiv papers from the astro-ph section. All knowledge sources are stored in ChromaDB vector storage for semantic retrieval. The central ReAct (Reasoning + Acting) agent orchestrates responses by exploring available knowledge sources through explicit reasoning steps, following a hierarchical approach that prioritizes course materials, then reference textbooks, and finally inherent LLM knowledge. A dedicated Moderator Agent ensures educational integrity by evaluating all responses against pedagogical guidelines, preventing direct homework solutions, and maintaining pedagogical teaching principles. The system uses Gemini as backbone with real-time monitoring through Papertrail logging, incorporating mechanisms to reduce hallucinations through trusted, domain-specific sources.}
    \label{fig:astrotutor_architecture}
\end{figure*}

\begin{itemize}
\item \textbf{RAG Course Material Agent}: The primary retrieval system searches through the lecturer's course materials (lectures, tutorials) adapted from the lecturer's textbook, now released as the online textbook ``Statistical Machine Learning for Astronomy.'' Having trusted, domain-specific sources is important for maintaining accuracy and preventing hallucinations that plague general-purpose models \citep{Alkaissi2023,Huang2023,Towhidul2024}. This agent employs Gemini's embedding-004 model\footnote{\url{https://ai.google.dev/gemini-api/docs/embeddings}} to create vector representations of both queries and knowledge base content, enabling semantic matching between student questions and relevant course materials.

\item \textbf{ReAct Agent}: The central orchestrator coordinates between different tools and agents based on student queries. This agent follows a ``Reasoning + Acting'' paradigm, first analyzing the student's question through explicit reasoning steps, determining which tools are needed, executing actions, observing results, and synthesizing responses from multiple sources. The ReAct framework \citep{Yao2022} allows iterative information gathering and ensures thorough responses by systematically exploring available knowledge sources.

\item \textbf{Reference Textbook Agent}: While course materials provide focused coverage of astrostatistics and machine learning techniques, students often require broader theoretical foundations extending beyond immediate class content. Additional RAG systems access Bishop's ``Pattern Recognition and Machine Learning'' \citep{Bishop2006} and other trusted textbooks for advanced theoretical explanations, providing deeper mathematical foundations when students require additional theoretical background.

\item \textbf{Moderator Agent}: A quality control system evaluates all generated responses against predefined pedagogical guidelines using a separate LLM call. The moderator ensures responses maintain appropriate academic standards, use correct mathematical formatting, avoid providing complete homework solutions, and maintain educational integrity by guiding rather than solving.
\end{itemize}

\begin{figure*}[ht]
    \centering
    \includegraphics[width=\textwidth]{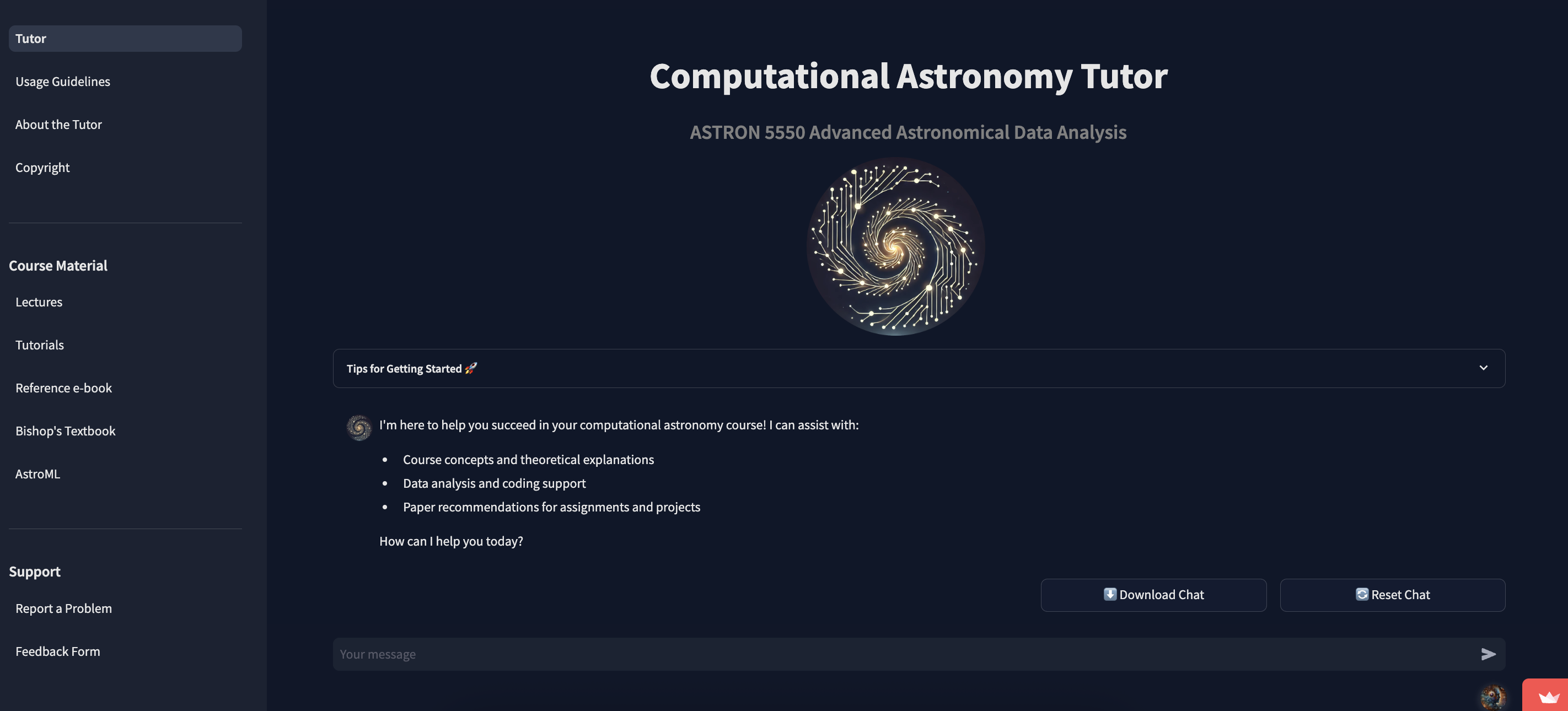}
    \caption{User interface of the AstroTutor system. The interface provides organized access to course materials including lectures, tutorials, and reference textbooks. The main chat interface facilitates pedagogical interactions, offering assistance with course concepts, data analysis and coding support, and paper recommendations for assignments and projects. Students can download chat histories and reset conversations as needed.}
    \label{fig:astrotutor_interface}
\end{figure*}

The system follows a hierarchical approach: first consulting course materials, then auxiliary textbooks, and only using inherent LLM knowledge when relevant information cannot be found in trusted sources. The system utilized a modular configuration approach with YAML files controlling model parameters and separate instruction files defining agent behavior. The ReAct agent was configured with a maximum of 20 iterations and verbose logging for debugging and analysis. Context management was handled through ChromaDB\footnote{\url{https://www.trychroma.com/}} persistent storage, with separate collections for different knowledge domains to prevent cross-contamination and optimize retrieval accuracy. The system included logging through Papertrail\footnote{\url{https://www.papertrail.com/}} for monitoring usage patterns and identifying potential issues in real-time.

\subsection{ArXiv Integration and Knowledge Base Expansion}

Traditional coding assessments face growing challenges from LLM capabilities \citep{Finnie-Ansley2022}, leading us to reimagine this upper-level astrostatistics and machine learning course to focus on authentic problem-solving. Previously, such courses typically provided students with predetermined project lists that relied heavily on the lecturer's expertise and limited students to familiar problems. We sought to extend beyond this limitation by enabling students to discover research directions independently.

\begin{itemize}
\item \textbf{Paper Recommendation Agent}: A specialized RAG system focused on a curated arXiv database, designed to suggest relevant research papers based on techniques students are studying or wish to apply in their projects. This agent maintains separate vector databases for different technique categories.
\end{itemize}

For supporting group projects, we expanded AstroTutor's knowledge base by using structured summaries and concept extraction from approximately 400,000 papers from the astro-ph section of arXiv \citep{Ting2025astromlab5}. This dataset provides summaries organized into semantic sections and identifies key methodological concepts from each publication. From these concepts, we curated technique-specific subsets containing 500-3,000 papers relevant to individual course techniques (such as linear regression, Principal Component Analysis, or Gaussian Mixture Models), creating vector embeddings using Gemini's text-embedding-004 model for semantic retrieval. 

The system's output format for paper recommendations included paper titles and authors extracted from summary documents, detailed descriptions of relevance to student queries, and direct arXiv links. For example, a PCA applications query would return a curated list of 10-15 papers with explanations of how each paper's methodology related to the student's specific question or project needs, organized by relevance and methodological similarity. This approach provided students with precise summaries and direct links to relevant papers without the hallucinations typically associated with general-purpose LLMs, serving as a research library for exploring potential project ideas through natural language interactions.

\section{LLM Usage}

Throughout the semester, students were informed about AstroTutor and its advantages and limitations, but were not required to use it exclusively. We emphasized the importance of students learning to critically evaluate various LLMs of their choice. Students were encouraged to document their usage of both AstroTutor and general-purpose LLMs for all course assignments through bonus questions at the end of each assignment, with bonus points awarded for thoughtful reflection on their AI interactions. Failed attempts and challenges were particularly welcomed in these reflections. From these self-reports we are able to understand their usage patterns, which are reported here.

We acknowledge that this incentive structure introduces potential biases in the self-reported data. Students may have been inclined toward reporting positive or constructive experiences to earn bonus points, and the non-anonymous nature of the reflections may have introduced social desirability bias. However, we note that bonus points were awarded for thoughtful engagement---including documentation of failures and frustrations---rather than for positive sentiment specifically. Moreover, the observed trend of decreased LLM reliance is arguably less susceptible to social desirability bias, as students might have assumed the instructor preferred to see increased engagement with the tools he developed and introduced. Nevertheless, the findings should be interpreted with these limitations in mind, and future studies should consider triangulating self-reports with objective performance-based measures.

\begin{figure}[ht]
    \centering
    \includegraphics[width=0.45\textwidth]{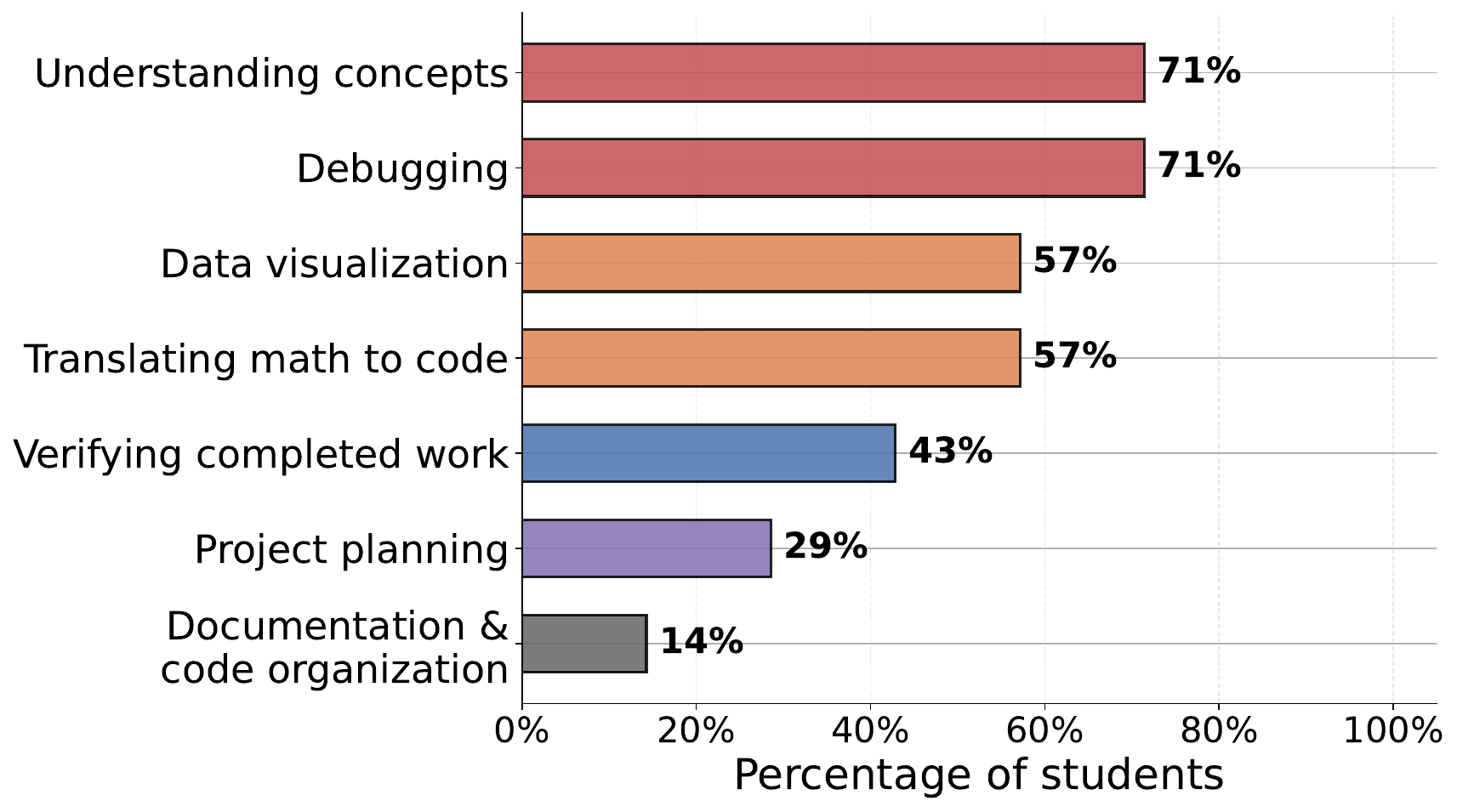}
    \caption{Distribution of LLM usage patterns among students based on post-course survey responses ($n=7$ respondents out of 12 enrolled). The horizontal bar chart displays the frequency of different task categories where students employed LLMs throughout the semester. Understanding concepts and debugging were the most common applications ($\sim$70\% each), reflecting students' primary use of AI tools for comprehension and technical problem-solving. Data visualization and translating mathematical concepts to code were moderately used ($\sim$60\% each), while verification of completed work showed intermediate adoption ($\sim$40\%). Project planning and documentation tasks showed lower but notable usage ($\sim$30\% and $\sim$10\% respectively).}
    \label{fig:usage_patterns}
\end{figure}

The qualitative themes reported in this paper were identified through an iterative process: the authors read all student reflections across the four homework assignments, and LLM-assisted text processing was used to organize and categorize the recurring patterns. All identified themes were subsequently verified by the authors against the original student responses. The complete survey items are presented in Figures~\ref{fig:usage_patterns}, \ref{fig:poll_results}, and \ref{fig:course_evaluation}.

Rather than restricting access or providing limited guidance, no specific instructions were provided on how to use these tools, allowing observation of natural patterns of adoption and adaptation. The diversity of tools used illustrates both the rapidly evolving landscape of LLM applications and students' willingness to explore multiple options to find those best suited to their specific needs.

\begin{figure}[ht]
    \centering
    \includegraphics[width=0.45\textwidth]{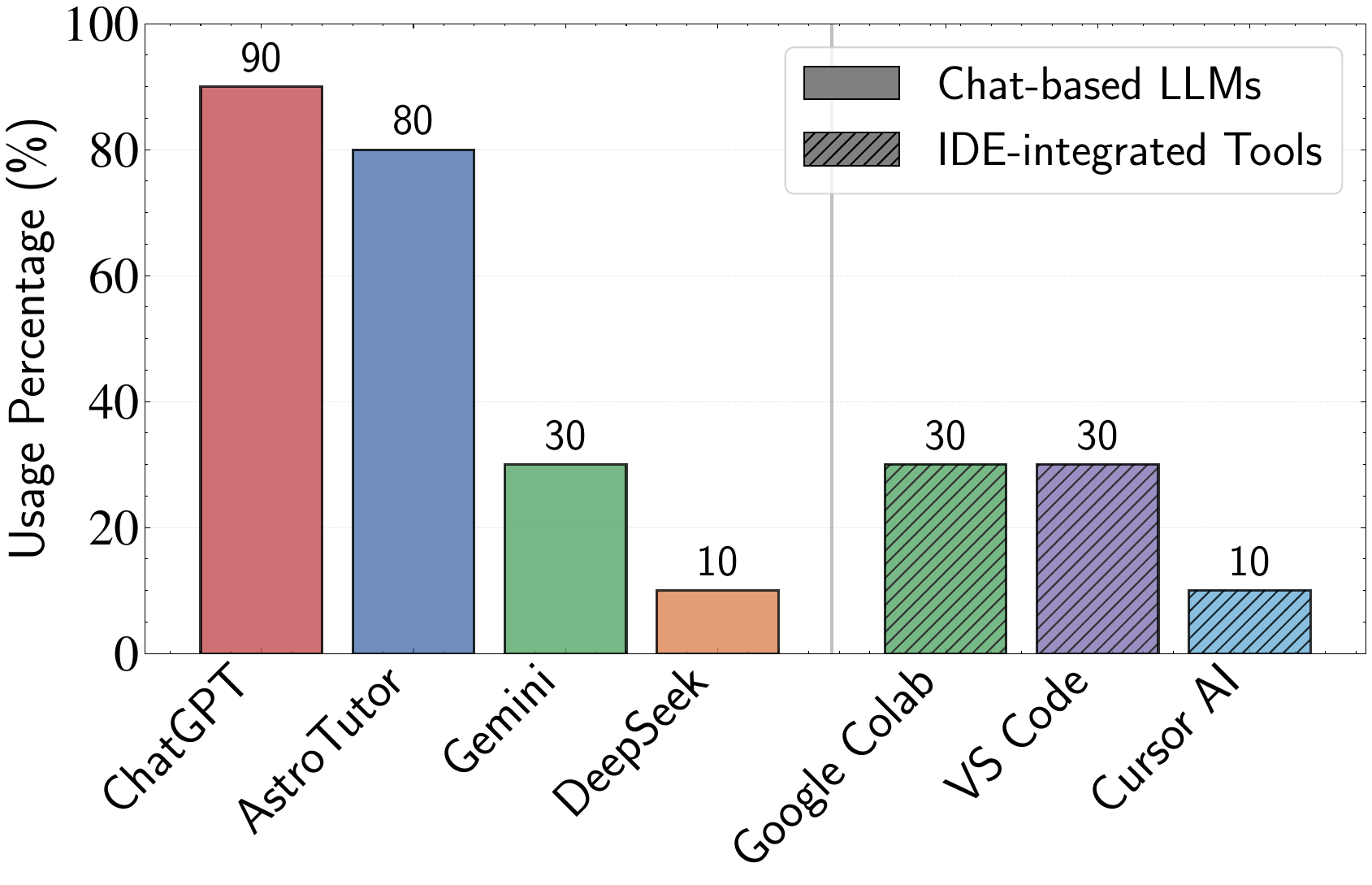}
    \caption{Distribution of LLM tool usage among students throughout the semester. The chart displays usage percentages for chat-based LLMs (solid bars) and IDE-integrated tools (hatched bars). ChatGPT was the dominant tool with 90\% adoption, followed by AstroTutor at 80\%. Students typically used AstroTutor for theoretical understanding and ChatGPT for coding assistance, demonstrating complementary roles. For IDE-integrated tools, Google Colab and Visual Studio Code (with GitHub Copilot) each achieved 30\% adoption, while Cursor AI reached 10\%. This indicates limited familiarity with modern IDEs among astronomy students, suggesting a gap in computational training that requires attention. Based on $n=10$ students who submitted homework reflections out of 12 enrolled. Usage data collected through self-reporting in assignment bonus questions.}
    \label{fig:llm_usage}
\end{figure}

\textbf{AstroTutor} was used by 80\% of reporting students, making it the second most popular tool after ChatGPT. Students appreciated AstroTutor's domain-specific knowledge but found different use cases for it compared to general-purpose LLMs. Many students developed a pattern of relying on AstroTutor for understanding course material and theoretical concepts while turning to ChatGPT primarily for coding assistance and implementation details. Students reported that AstroTutor produced fewer hallucinations than general-purpose models, likely due to its reliance on curated, domain-specific sources through RAG, but students were still inclined to use general models intermittently because they often provided more direct responses. The pedagogical approach with lengthy, questioning responses from AstroTutor might have discouraged some students, highlighting an ongoing challenge in balancing educational depth with user preference for directness. 

This tension between Socratic guided inquiry and student preferences for efficiency is not unique to our implementation---\citet{Tufino2025} reported a similar finding with their RAG-based Socratic physics tutor, where some users became frustrated when the system persistently withheld direct solutions. The challenge is fundamental: students working under time pressure on assignments naturally gravitate toward tools that provide immediate answers, even when guided discovery may produce deeper learning. This creates a risk that pedagogically principled AI tutors may inadvertently drive students toward less structured commercial alternatives. Possible mitigation strategies include adaptive scaffolding---where the tutor progressively provides more direct guidance after detecting repeated student difficulty---or hybrid modes that allow students to toggle between Socratic and direct response styles depending on the nature of their query.

\textbf{ChatGPT}\footnote{\url{https://platform.openai.com}} emerged as the dominant tool, used by 90\% of reporting students. This widespread usage demonstrates how critical this study is—turning a blind eye to AI usage is not a viable option. Students referenced various models including standard ChatGPT (GPT-4o) and reasoning models (o1, o3). Many students demonstrated awareness of different capabilities across model versions, with several specifically mentioning using the reasoning-enhanced versions for more complex tasks.

\textbf{Google's Gemini} was used by about 30\% of reporting students, particularly in conjunction with Google Colab. Since students were already familiar with Google Colab from previous courses, integrating Gemini represented a minimal transition. Students appreciated the integrated AI assistant functionality within the Google Colab environment.

\textbf{DeepSeek}\footnote{\url{https://platform.deepseek.com/}} (including v3 \citep{Deepseek2024} and R1 models \citep{Deepseek2025}) was used by 10\% of reporting students, indicating that students were generally aware of cutting-edge tools. Some students encountered access limitations, with several reporting they needed to switch from DeepSeek to alternative LLMs due to server capacity issues, especially during high-demand periods in early 2025.

One notable absence was \textbf{Claude} \footnote{\url{https://console.anthropic.com/}}, which is arguably one of the leading performers in coding assistance \citep{Kevian2024}\footnote{\url{https://www.swebench.com/}}. This highlights that even in a major astronomy program, tool selection often comes from somewhat random information consumption rather than systematic evaluation. While students demonstrated the ability to make reasonable choices, some guided introduction to available tools might have been beneficial, as students may not be aware of all quality options available to them.

\subsection{IDE-Integrated Tools}

Integrated Development Environments (IDEs) are software applications that provide facilities for software development, typically including code editors, debuggers, and build automation tools. We consider integration of LLMs in IDEs to be a critical skill for modern astronomers, as these tools provide more seamless workflows by offering contextual assistance directly within coding environments.

We briefly mentioned IDEs at the beginning of the course, and students observed the lecturer ``live coding'' with Cursor during most tutorials, seeing firsthand how an expert interacts with IDEs to speed up productivity. Most students reported that seeing the lecturer live-coding during lectures was particularly eye-opening. However, students generally showed limited familiarity with modern IDE environments, even within a major astronomy program. This indicates a gap in computational training within the astronomy community that could benefit from more explicit instruction.

\textbf{Google Colab} with its integrated AI assistant (Gemini) was mentioned by about 30\% of students, providing seamless integration with their existing workflow. Students were already familiar with Google Colab from previous courses, making this integration natural.

\textbf{Visual Studio Code}\footnote{\url{https://code.visualstudio.com/}} with GitHub Copilot\footnote{\url{https://copilot.github.com/}} was mentioned by about 30\% of students. Students reported that GitHub Copilot enhanced their coding efficiency by providing contextual suggestions based on existing notebook content, helping them answer questions more rapidly.

\textbf{Cursor AI}\footnote{\url{https://www.cursor.com}} was mentioned by 10\% of students. Cursor has risen considerably in the computer science community. Interestingly, no students knew about this modern tool until it was mentioned in a tutorial session where the lecturer demonstrated it live. Students who adopted it appreciated the in-line generator and chat features, noting how the integrated nature of the tool streamlined their workflow by maintaining awareness of the full coding context.

However, despite seeing it demonstrated live, the subscription costs compared to more readily available options like Google Colab and GitHub Copilot appeared to be a limiting factor, highlighting the importance of resource accessibility. While subscription costs may appear minimal for professional researchers, at the student level they can still constitute a barrier. We thus suggest that institutions or departments invest in such tools, just as most departments now invest in Overleaf and Slack, and recognize their importance. Hopefully this paper will encourage movement in that direction.

\vspace{2cm}
\section{Common Usage Patterns}

Through students' self-reporting in homework assignments and post-course surveys, we observed seven main usage patterns for LLMs (Figure~\ref{fig:usage_patterns}). The frequency reveals that students primarily used LLMs for conceptual understanding and technical problem-solving rather than task completion shortcuts.

\textbf{Understanding Concepts.} Students frequently used LLMs to clarify concepts and problem statements, making this the most common application (70\% usage). This included understanding mathematical formulas, clarifying assignment questions, and connecting theoretical concepts to practical implementations. Since most assignments involved applying techniques to real astronomical research problems, students with different background levels needed varying degrees of context to tackle questions (for example, understanding what Baryon Acoustic Oscillations are). The high usage frequency suggests that students valued LLMs primarily as learning tools for comprehension.

\textbf{Debugging.} Debugging was equally dominant (70\% usage), going beyond simple syntax errors to include conceptual issues in algorithm implementation. Matrix operations proved particularly challenging due to their notational subtleties—issues with transposition, dimension matching, broadcasting, and proper matrix multiplication sequence caused difficulties. These linear algebra issues were particularly problematic when working with high-dimensional astronomical data, where matrix shape mismatches could be difficult to diagnose but catastrophic for results. Numerical stability issues presented another challenge where LLMs provided valuable assistance. Common issues included NaN values appearing in loss calculations, overflow errors in exponential functions, and instability in normalization procedures. Students learned important numerical computing principles through these interactions, such as the technique of subtracting the largest logit value when implementing softmax functions to prevent exponential overflow.

\textbf{Translating Math to Code.} Converting mathematical formulas to code was moderately common (60\% usage). These tasks often involve implementation details that can be time-consuming but don't necessarily deepen conceptual understanding. Students sought help translating their understanding of assignment questions into functional code, generating implementations for specific algorithms, or explaining code structure for complex operations. This was particularly evident for algorithms like Principal Component Analysis, Gaussian Mixture Models, and logistic regression. The course encouraged students to build these methods from scratch rather than simply calling functions from libraries like scipy or astropy. Students reported that their experience with manual implementations from tutorials helped them craft better prompts when asking about built-in functions.

\textbf{Data Visualization.} Creating visualizations was another moderate application (60\% usage). Students sought help with advanced techniques like overlaying images on plots, creating 2D density plots, and implementing specialized color maps appropriate for astronomical data. A common visualization challenge involved overlaying images on plots, particularly for tasks like placing thumbnail images on UMAP 2D embeddings to visualize clustering results. Students also requested guidance on creating multi-panel visualizations with subplots, adding custom annotations and labels, and incorporating inset plots to highlight specific features. Students found that LLMs were particularly helpful for these visualization tasks, as they could provide sample code that demonstrated both basic implementation and customization options that improved the clarity and professional appearance of their results.

\textbf{Verifying Completed Work.} A portion of students developed verification practices (40\% usage). After completing problems independently, students presented solutions to LLMs for correctness checks and improvement suggestions. This represented an evolution in AI integration—students leveraged LLMs as validation tools after applying their own understanding rather than for initial solution generation. Students reported this practice helped them gain confidence, identify misunderstandings, and discover alternative approaches.

\textbf{Project Planning.} Moderate usage occurred for project-related tasks (30\%). Open-ended group projects required students to formulate research questions from literature, identify data sources, and develop methodologies. Students employed AstroTutor's curated literature database (500-3,000 arXiv papers per topic) to discover research directions through natural language queries. LLMs helped identify and navigate specialized astronomical databases, with domain knowledge acquisition becoming important for learning concepts like the Tully-Fisher relation or advanced techniques like No U-Turn Sampler for MCMC algorithms.

\textbf{Documentation and Code Organization.} The least common usage involved documentation tasks (10\%). Students requested help generating meaningful comments for complex algorithms and creating structured documentation, improving readability and developing better coding practices. LLMs also assisted with parsing unusual astronomical data formats and technical writing tasks like creating properly formatted references.

\begin{figure*}[ht!]
    \centering
    \includegraphics[width=\textwidth]{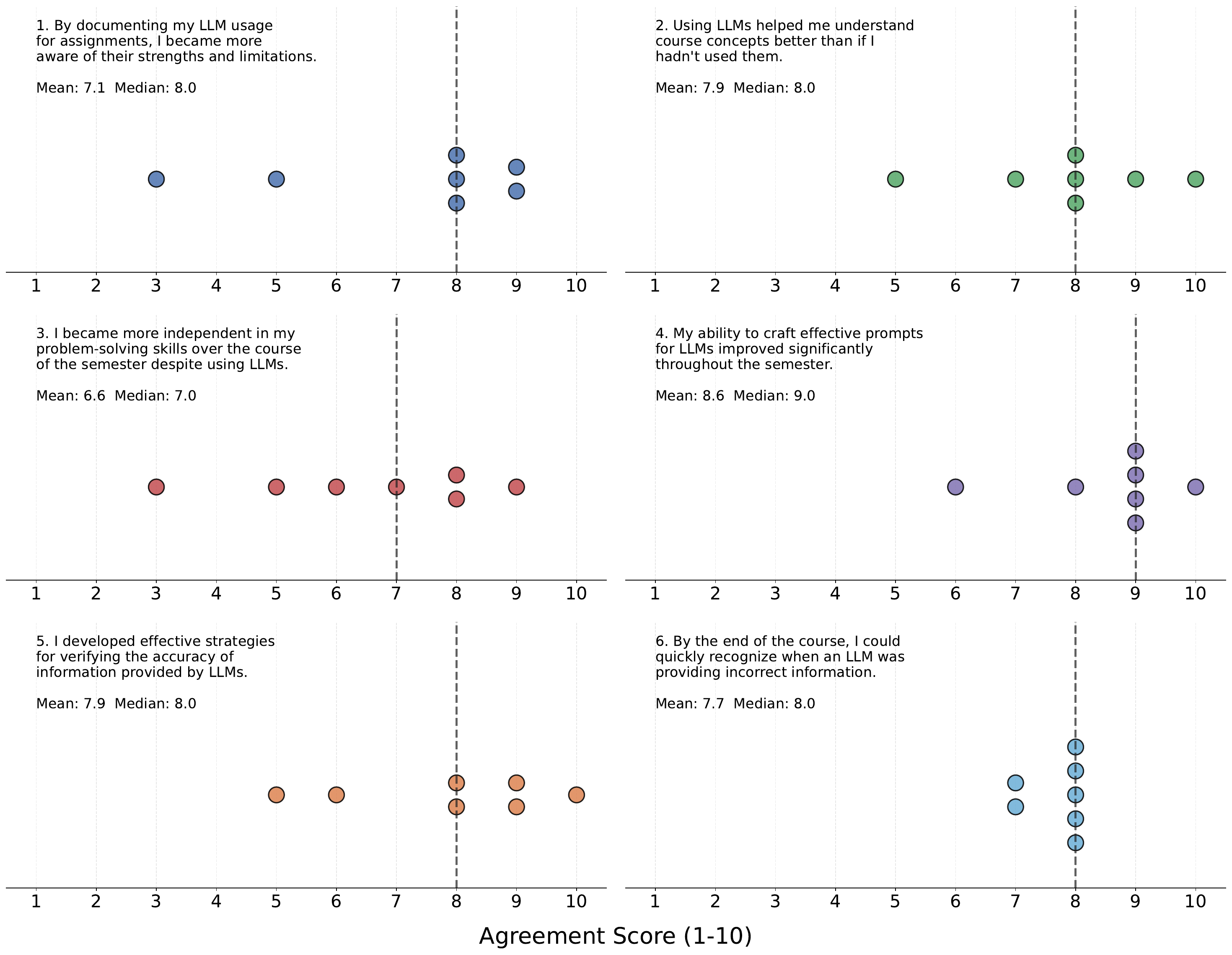}
    \caption{Student self-assessment of learning outcomes and LLM proficiency development across six key dimensions (1-10 Likert scale) based on post-course survey responses ($n=7$ respondents out of 12 enrolled). Each panel shows individual student responses as dots (stacked when multiple students gave the same score), with a dashed vertical line indicating the median. The survey assessed: (1) awareness of LLM strengths and limitations through documentation, (2) LLM effectiveness for concept understanding, (3) maintenance of problem-solving independence despite LLM use, (4) improvement in prompt crafting abilities, (5) development of verification strategies for AI-generated information, and (6) ability to recognize incorrect LLM outputs. Mean and median scores are displayed for each dimension. The results reflect student perceptions of their learning progression and AI literacy development throughout the semester.}
    \label{fig:poll_results}
\end{figure*}

\begin{figure*}[ht!]
    \centering
    \includegraphics[width=\textwidth]{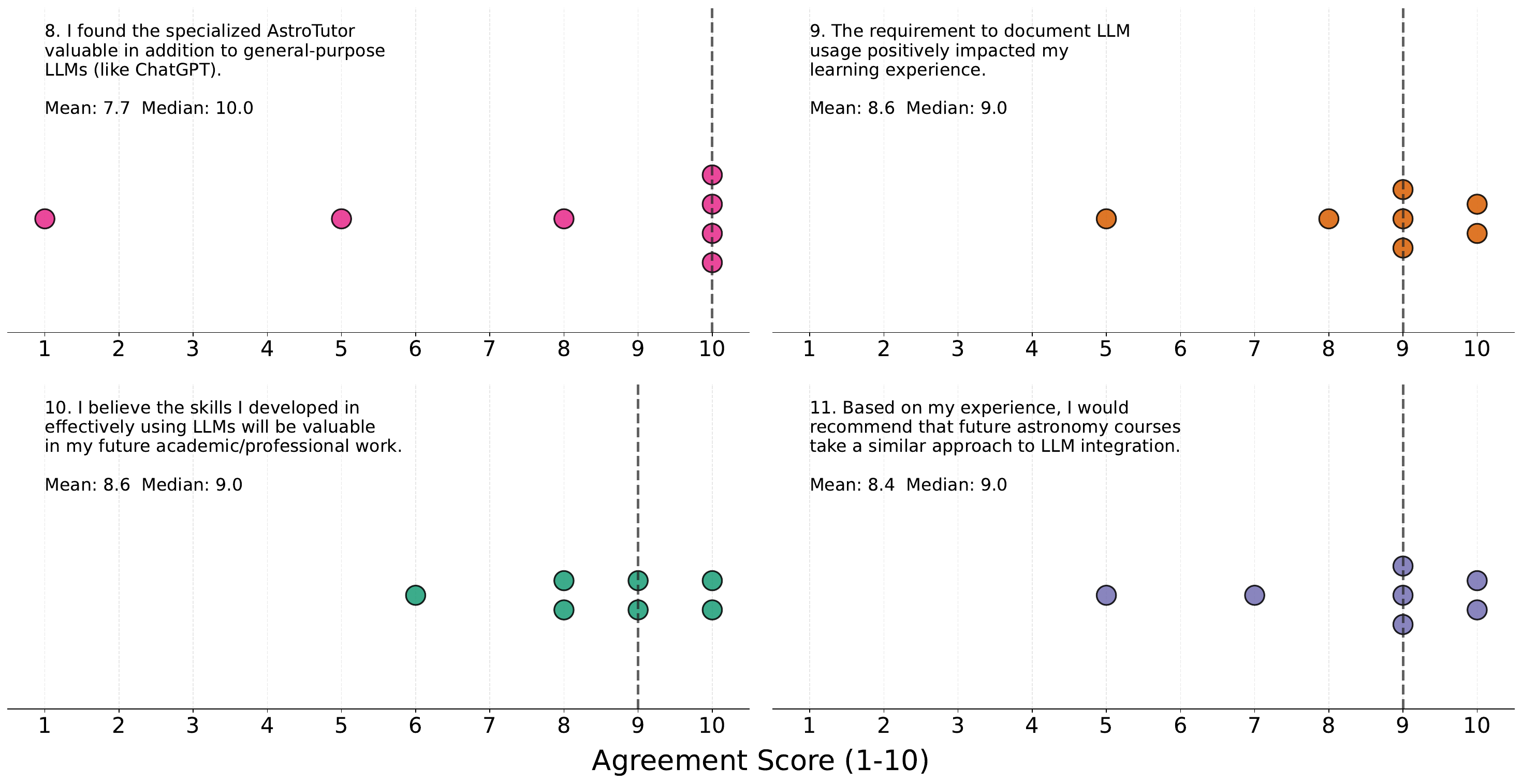}
    \caption{Student evaluation of course implementation and future implications (1-10 Likert scale) based on post-course survey responses ($n=7$ respondents out of 12 enrolled). Each panel shows individual student responses as dots (stacked when multiple students gave the same score), with a dashed vertical line indicating the median. The survey assessed: (8) perceived value of the specialized AstroTutor system compared to general-purpose LLMs, (9) impact of documentation requirements on learning experience, (10) anticipated value of developed LLM skills for future academic and professional contexts, and (11) recommendation for similar LLM integration approaches in future astronomy courses. Mean and median scores are displayed for each dimension. The results reflect student assessment of the educational intervention's effectiveness and its broader applicability in astronomy education.}
    \label{fig:course_evaluation}
\end{figure*}

\section{Evolution of LLM Usage}

We were interested in understanding whether students developed useful skills through LLM use. Given that implementing LLMs in daily problem-solving seems inevitable, we chose to encourage their use—with critical thinking—rather than banning them. These findings are based on students' self-reporting throughout the semester and confirmed by post-course survey results (Figures~\ref{fig:poll_results} and \ref{fig:course_evaluation}).

The question was whether actively encouraging LLM use would improve students' skills or worsen educational outcomes. Based on student self-reports and survey data, the evidence suggests the former, with \textbf{our key finding being that students reported decreased dependence on LLMs over time rather than increasing it}—contrary to common assumptions that encouraging LLM use will worsen learning outcomes. Students also developed a community of practice around effective LLM use, with evidence of informal strategy sharing that enhanced their AI proficiency skills. Post-course surveys confirm that students maintained problem-solving independence despite LLM use (median score: 7.0), and that the documentation requirement positively impacted their learning experience (median score: 9.0). 

We note that the independence dimension (Q3) received the lowest median score among all surveyed items, reflecting the fact that while students reported decreased LLM usage over time, perceived independence is multidimensional---some students may have felt that their problem-solving still benefited substantially from LLM availability, even as they used these tools less frequently and more strategically.

\textbf{Early Assignment Patterns.} In early assignments, students primarily used LLMs for basic clarification and simple coding help. Their initial prompting strategies were underdeveloped, characterized by casual, conversational approaches without attention to prompt engineering \citep{Chen2024,Knoth2024,Schulhoff2024}. Students would write general requests for explanations or ask broad questions without providing context or structure. Students often blamed themselves for miscommunications with AI during this early phase.

As the semester progressed, students developed more sophisticated approaches to AI interaction, moving beyond basic assistance-seeking to strategic engagement patterns.

\textbf{Domain-Specific Role Prompting.} Several students independently discovered the effectiveness of domain-specific role prompting \citep{Xu2023,Zheng2023,ChenJ2024,Kong2024}, instructing models like ChatGPT to assume the persona of an expert astronomer with specialization in machine learning. This approach helped focus the model on relevant domain knowledge and produced more field-appropriate responses. This evolution demonstrates students developing meta-learning skills alongside course content. Students reported notable improvement in their ability to craft effective prompts throughout the semester (median score: 9.0), the highest score among all assessed dimensions. 

\textbf{Targeted Questioning and contextual enrichment.} Instead of asking LLMs for complete solutions, students learned to ask about specific concepts. They discovered that prompts focused on discrete topics were more effective than copying and pasting entire assignment questions. This development was partly by design—anticipating LLM use, some assignment questions were deliberately less structured, avoiding breaking equations into fine-grained steps that would be easy to copy and paste. Students demonstrated growth in learning how to break down questions into manageable pieces that LLMs could handle effectively. This problem structure also helped students develop better decomposition skills. Their prompting strategies evolved to include contextual enrichment: providing code snippets for debugging, offering problem context before asking specific questions, and including previous problem statements to ensure continuity. Students discovered that sharing entire notebooks or code sections produced more accurate and helpful responses compared to sharing isolated fragments, as complete context allowed the AI to better understand dependencies, variable definitions, and overall structure. 

\textbf{Enhanced Learning Focus.} Students valued AI's ability to bridge the gap between theoretical concepts from lectures and their practical implementation in assignments, with this theory-to-practice translation being helpful for connecting abstract mathematical concepts to concrete computational implementations. Students appreciated the rapid implementation assistance and coding efficiency, which allowed them to focus more on understanding concepts rather than getting stuck on implementation details. This time efficiency enabled students to concentrate on conceptual learning rather than minutiae of coding syntax and technical implementation issues. Students with different learning approaches benefited differently—visual learners appreciated assistance with visualization code, while others found conceptual explanations most valuable. Survey results support this observation, with students reporting that LLMs helped them understand course concepts better than traditional methods alone (median score: 8.0).

\textbf{Verification Rather Than Generation.} Students autonomously began asking AI to review their completed work rather than generate solutions from scratch. They would complete solutions independently and then submit their work to the LLM for evaluation, verification, and suggestions for improvement. This approach allowed students to maintain their learning autonomy while benefiting from AI assistance. Students developed iterative refinement strategies: breaking complex problems into smaller steps addressed sequentially, building sequences of related prompts to maintain context, cross-checking outputs against course materials, and trying the same prompt with different tools when facing unclear explanations. Post-course assessments confirm that students developed effective strategies for verifying AI-generated information (median score: 8.0).

\textbf{Critical Tool Comparison.} By experimenting with multiple tools, students developed the ability to evaluate outputs and recognize each tool's strengths and weaknesses. A common pattern emerged where students would use general-purpose models like ChatGPT primarily for coding assistance while relying on AstroTutor for theoretical explanations. Students found the specialized AstroTutor valuable as a complement to general-purpose LLMs (median score: 10.0), though with notable variation in responses (standard deviation: 3.2). In later assignments, students became more sophisticated in recognizing when the AI was at fault rather than blaming themselves. Students described becoming more confident in their AI interactions over time, developing strategies to cope with inconsistent quality in AI responses, such as rewording the same question multiple times or triangulating between different AI tools. By course completion, students reported confidence in recognizing incorrect LLM outputs (median score: 8.0).

\textbf{Decreased Dependence.} This represents our key observational finding: instead of increasing LLM usage, actively asking students to document and think critically about their usage appeared to lead many to rely on LLMs progressively less as the course advanced. Students reported developing better debugging skills and greater ability to resolve simpler issues independently. Multiple students reflected on the trade-off between quick AI solutions and deeper learning, showing awareness of the potential to use AI as either a shortcut or a learning enhancer. This pattern suggests that LLMs were serving as effective learning scaffolds rather than creating permanent dependencies.

The growing independence stems from students having developed awareness of LLM limitations—recognizing different error rates between models, understanding the overly affirming nature of LLMs that can impede learning, and developing habits of cross-checking AI outputs against authoritative sources. Documentation requirements contributed to this awareness development (median score: 8.0). This suggests that encouraging active documentation of LLM use, combined with critical reflection, is a key approach to fostering independence rather than dependence in AI-assisted learning. Students strongly believe the skills they developed will be valuable for future academic and professional work (median score: 9.0) and would recommend this approach for future astronomy courses (median score: 9.0).

\section{LLM-Specific Challenges}

The decreased reliance on LLMs observed throughout the semester stemmed partly from students identifying these challenges through active reflection encouraged by bonus points for documenting their AI interactions. Students showed growing understanding of tool limitations.

\textbf{Prompt Engineering Difficulties.} Students struggled with balancing prompt specificity — too general prompts yielded unhelpful generic answers, while overly specific prompts led to irrelevant details or misunderstood questions \citep{Denny2023,Babe2024}. Incorrect prompting could lead to an overcomplicated solution when a simpler approach was needed. When providing only code blocks and asking about function usage, AI would incorrectly assume existing code was correct. LLMs would sometimes generate completely unrelated responses when full assignment context was not provided, making problematic assumptions without complete information. Limited context windows and memory limitations required reestablishing context when switching models, disrupting natural problem-solving flow.

\textbf{Over-reaffirmation.} LLMs occasionally provided incorrect information with high confidence \citep{Peters2025}, particularly problematic for educational applications. Students reported confusion from misleading information that reinforced rather than corrected misconceptions. More fundamentally, LLMs are reluctant to directly contradict user statements or point out errors, instead offering gentle suggestions or building upon flawed premises. Students found this excessive positive reinforcement \citep{Malmqvist2024,Ruwe2024} frustrating compared to human instruction, where clear error correction plays a vital role in deepening understanding. Students noted that specialized tools like AstroTutor were more reliable than general-purpose models, showing awareness that different models have different strengths and limitations.

\textbf{IDE Integration Issues.} GitHub Copilot and similar tools would fixate on previous code patterns, over-training on bootstrapping code and requiring memory resets \citep{Barke2022}. This memory cache behavior compounds over-reaffirmation problems, as AI becomes increasingly confident in potentially incorrect patterns from previous interactions. These issues highlight the need for deeper understanding of IDE features and caching behavior in AI-assisted coding environments.

Many difficulties could be addressed through API usage, which remains largely unknown to students. Even at the graduate level, misconceptions persist that LLMs are only accessible through graphical interfaces, despite APIs offering more powerful interaction possibilities.

\begin{figure*}[htbp]
    \centering
    \includegraphics[width=1.0\textwidth]{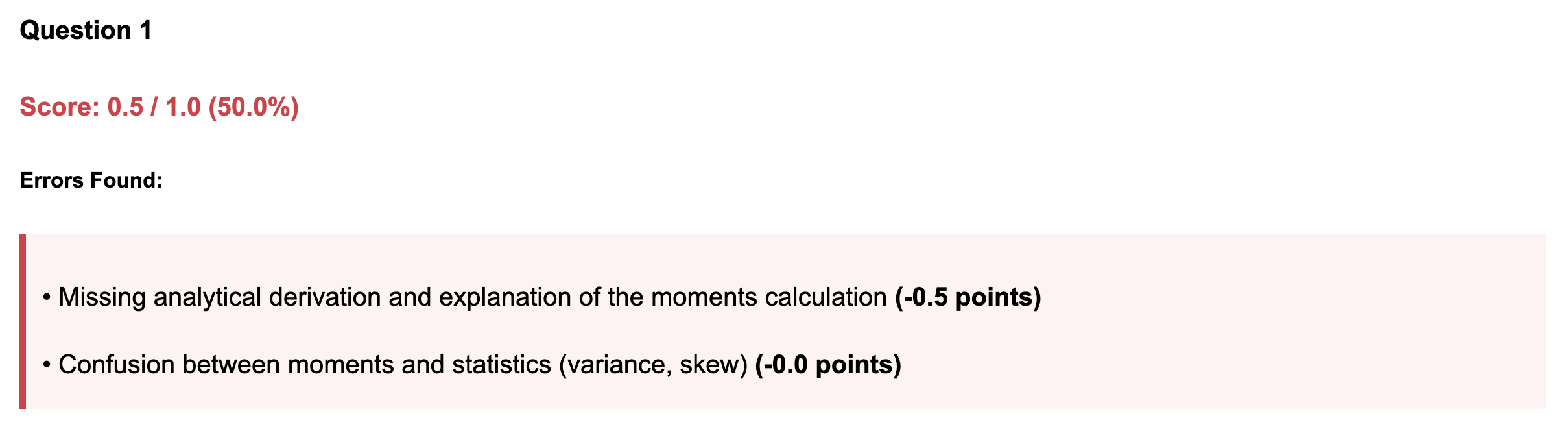}\\
    \includegraphics[width=1.0\textwidth]{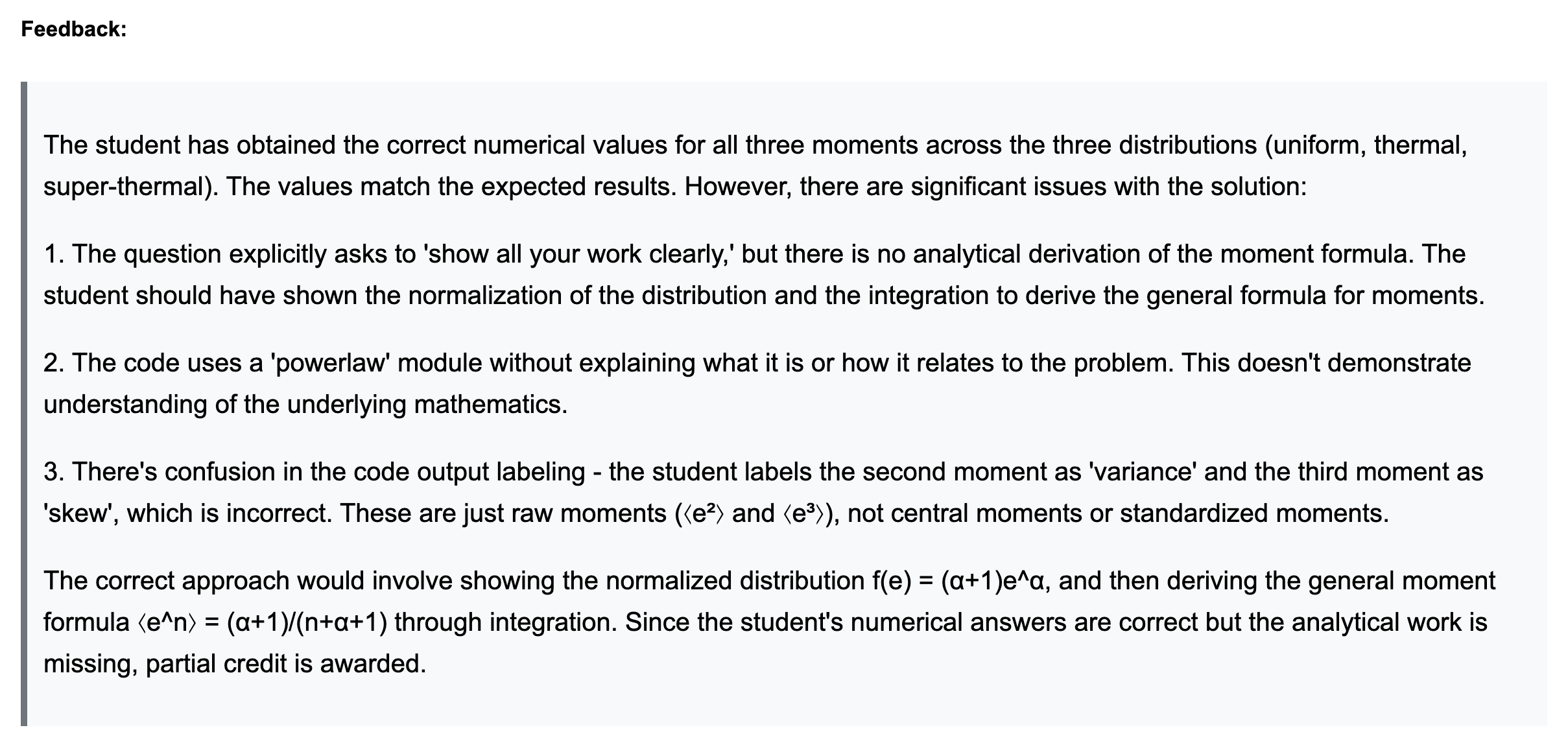}\\
    \caption{Example of LLM-generated grading feedback showing detailed error identification and constructive guidance for a student's analytical approach to calculating distribution moments.}
    \label{fig:llm_grading_example}
\end{figure*}

\begin{figure*}[htbp]
    \centering
    \includegraphics[width=1.0\textwidth]{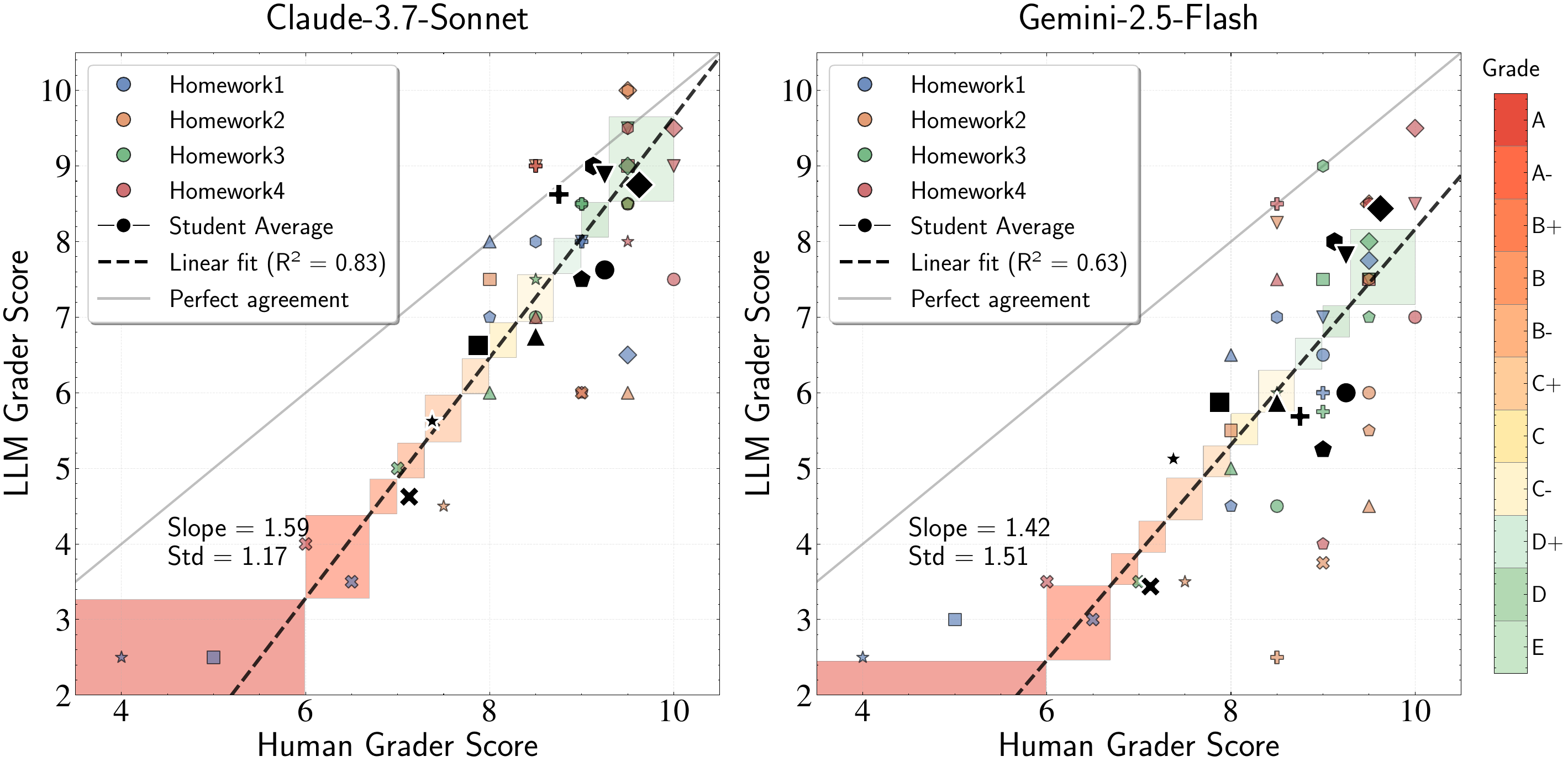}
    \caption{Comparison of LLM-assisted grading versus human grader scores across four homework assignments for two different models: Claude-3.7-Sonnet (left) and Gemini-2.5-Flash (right). The analysis includes $n=10$ students, each graded on four homework assignments (40 individual scores total). Individual homework scores are shown as colored points, with black circles representing student averages used for linear regression analysis. The dashed line shows the linear fit based on student averages, while the solid gray line indicates perfect agreement between grading methods. Colored grade regions represent bias-corrected grade boundaries (from E to A), where each region's boundaries are mapped through the regression line to account for systematic scoring differences. Claude demonstrates stronger correlation ($R^2 = 0.83$) and lower variability (Std = 1.17) compared to Gemini ($R^2 = 0.63$, Std = 1.51), though both models show systematic bias with slopes of 1.59 and 1.42 respectively, indicating that both LLM graders tend to be more stringent than human grading. Points falling within the same colored region would receive equivalent letter grades after correcting for systematic bias. Each student is represented by a different marker shape to distinguish individual performance patterns across assignments.}
    \label{fig:grading_comparison}
\end{figure*}

\section{Grading and Assessments with LLM}

Beyond chat-based AI assistance for student learning, homework feedback represents a critical component of undergraduate education. Recent work in physics education has begun exploring AI-assisted grading, including studies on grading handwritten thermodynamics exams \citep{Kortemeyer2024grading} and psychometric approaches for assessing confidence in AI-generated grades \citep{Kortemeyer2025psychometrics}. Traditional grading approaches face several challenges at the undergraduate level that LLMs could potentially address.

\textbf{Quality and Consistency Issues.} Teaching assistants are often slightly more senior students who may have taken different versions of the course. In rapidly evolving fields like data analysis and machine learning in astronomy, techniques can change between course iterations, leading teaching assistants to miss important details or apply outdated approaches. Large classes typically employ multiple graders with varying expertise levels, making uniform grading standards difficult to maintain across all submissions \citep{Mutambuki2018,Tong2020}.

\textbf{Timeliness and Feedback Depth.} Student teaching assistants, burdened with their own coursework and research responsibilities, often require time for grading, resulting in delayed feedback when it would be most educationally valuable. Time constraints frequently lead to terse comments that lack the detailed explanations students need for improvement. In our experience, human grader comments were typically two to three words (e.g., ``wrong calculation'' or ``missing plot''), whereas the LLM grading system produced multi-sentence feedback identifying the specific error, explaining why the approach was incorrect, and suggesting how to fix it (see Figure~\ref{fig:llm_grading_example} for an example of LLM-generated feedback). This contrast is partly attributable to the fact that human graders in our course were senior undergraduates who had only recently taken the course themselves, and who were balancing grading duties with their own academic workload. The situation becomes further complicated by late submission policies—even a single student requesting deadline extensions can delay solution release for the entire class, preventing timely learning reinforcement for students who submitted on schedule \citep{Hattie2007,Hounsell2008}.

LLMs offer potential solutions to these challenges through automated grading with immediate, detailed feedback that maintains consistency across all submissions. However, we emphasize that this should not be framed as replacing teaching assistants. TA-ships serve a crucial role as training grounds for future faculty, developing pedagogical skills, mentoring abilities, and disciplinary communication that are essential for academic careers. Rather than eliminating TA positions, AI-assisted grading could free TAs from routine grading tasks, allowing them to focus on higher-order pedagogical activities such as one-on-one mentoring, leading discussion sections, providing conceptual guidance, and developing their own teaching skills. In this view, the TA role evolves from primarily a grader to primarily a mentor and pedagogical apprentice, which arguably better serves both the students and the TAs' professional development. We experimentally tested LLM grading alongside traditional human grading throughout this course. However, given ongoing ethical considerations that require broader community consensus \citep{Holmes2021,Akgun2022,KumarR2023}, all official grades released to students remained based on human evaluation. We conducted LLM grading as a parallel experiment and report our comparative findings here.

\subsection{Grading Implementation}

The course included four homework assignments, each consisting of approximately 8 coding questions built around simplified research papers. The lecturer deconstructed published research, extracted relevant datasets, and created step-by-step coding exercises that allowed students to reproduce key findings. The four homework topics were: (a) solving binary ellipticity from phase-space observations using Gaia data through Bayesian inference, (b) calibrating telescope responses with linear regression, (c) identifying strong gravitational lenses in James Webb Space Telescope data using logistic regression, and (d) detecting quasar outliers from spectra through Principal Component Analysis and Gaussian Mixture Models.

We developed an automated grading system using two LLM models: Claude-3.7-Sonnet, chosen for its coding reasoning capabilities, and Gemini-2.5-Flash, the more cost-effective option used in AstroTutor. The grading process involved parsing Jupyter notebooks to extract sequential question content, student code, outputs, and generated figures with automatic base64 image decoding and an organized directory structure. Questions were graded sequentially, with the LLM being provided full context of the current problem, the lecturer's solution, and student solutions from earlier questions to ensure consistent evaluation and avoid double-deducting points for recurring conceptual misunderstandings that might propagate through related problems.

The LLM grading system was designed to mirror effective human grading practices through carefully engineered prompts that followed the same grading philosophy communicated to human tutors. For each question, the LLM received the question content, the lecturer's reference solution, and the student's answer as a combined document. The grading prompt instructed the model to start from full marks, apply generous partial credit for correct approaches with minor errors, and accept alternative valid methods. Four evaluation dimensions were specified: understanding of core concepts, reasonable code implementation, presence of required figures and plots, and appropriate interpretation of results. The system generated structured JSON responses with specific fields including earned points, detailed error descriptions with point deductions, and feedback for each question. All scores were rounded to the nearest 0.5 points for consistency with human grading practices. The system generated detailed HTML reports with visual previews of student-generated figures and organized breakdowns by student and question. Figure~\ref{fig:llm_grading_example} shows an example of the detailed feedback provided by the LLM grading system, demonstrating both the identification of specific errors and constructive guidance for improvement.

\subsection{LLM Grading Performance Analysis}

The grading results demonstrate promising performance for LLMs as homework graders, particularly for stronger models like Claude. From a practical standpoint, Claude-3.7-Sonnet costs approximately USD \$0.10 per student homework grading, while Gemini-2.5-Flash operates at roughly five-fold lower cost. However, both models exhibit systematic differences from human grading patterns, with regression slopes consistently greater than one, indicating that LLM grading tends to be more stringent than human evaluation. This is visually apparent in Figure~\ref{fig:grading_comparison}, where the regression line sits below the perfect-agreement diagonal across most of the observed score range, meaning the LLM assigns lower scores than the human grader for the majority of students.

Despite this systematic bias, the correlation analysis reveals strong consistency in relative student rankings. We base the regression on student-average scores (averaged across four homework assignments) because the practical question for educational fairness is whether LLM grading would change a student's final letter grade — and letter grades are determined by semester averages, not individual assignment scores. When analyzing average homework scores for individual students and fitting a linear regression model, Claude achieves $R^2 = 0.83$ with human graders, meaning that student performance rankings are largely preserved—arguably the most critical aspect for educational fairness. Figure~\ref{fig:grading_comparison} illustrates this relationship, showing that while absolute scores differ, the rank ordering remains robust across the grade distribution. The bias-corrected grade boundaries demonstrate that students would generally receive equivalent letter grades after accounting for the systematic scoring difference.

For more economical models like Gemini Flash, performance degradation becomes apparent with $R^2 = 0.63$ and higher variability. While the regression slope is fitted to student averages, the standard deviation is calculated from residuals of all individual homework scores relative to this student-level regression line. This yields standard deviations of 1.51 for Gemini Flash versus 1.17 for Claude, indicating that Gemini shows greater inconsistency at the individual assignment level. While Gemini Flash still serves as a reasonable study companion, providing structured feedback and error identification, the reduced grading consistency limits its utility for formal assessment. The scatter in individual homework scores reveals that Gemini's performance varies across different question types and complexity levels, making it less reliable as a homework grader compared to Claude.

Given the systematic stringency observed in LLM grading, a critical question emerges: are human graders missing legitimate errors, or are LLMs identifying phantom problems? The extensive documentation of errors from both LLM and human graders enables detailed question-by-question comparison, revealing that the former explanation predominates.

\subsubsection{Comparing Claude-3.7-Sonnet with Human Grading}

\textbf{Computational precision.} Claude demonstrates superior performance in identifying technical implementation errors that human graders frequently overlook. These include mistakes in matrix transformations, gradient calculations, and normalization procedures in machine learning contexts. For instance, Claude consistently caught transposition errors in gradient computations where students incorrectly implemented matrix operations, while human graders often missed these errors that fundamentally affect algorithmic correctness. Such technical subtleties prove challenging for human graders to detect consistently across large numbers of submissions.

\textbf{Systematic consistency.} Claude exhibits uniformity in maintaining evaluation standards across students. Analysis of scoring patterns reveals that human graders showed variance in weighting similar types of errors across different students, while Claude applied consistent criteria throughout. Human graders demonstrated inconsistent penalization for equivalent conceptual omissions—such as incomplete Bayesian implementations or missing mathematical derivations—whereas Claude maintained uniform scoring based on specific technical deficiencies.

\subsubsection{Comparing Claude with Gemini-2.5-Flash}

\textbf{Contextual focus and prioritization.} The performance gap between Claude-3.7-Sonnet and Gemini-2.5-Flash primarily manifests in appropriate error identification. Claude correctly identifies complex technical errors while providing specific, actionable feedback, whereas Gemini Flash often over-penalizes superficial issues due to limited context memory. Gemini Flash sometimes makes inappropriate logical judgments, such as heavily penalizing students for acceptable visualization design choices when pedagogical objectives were clearly met, demonstrating a failure to distinguish between stylistic preferences and fundamental errors.

\textbf{Systematic consistency.} Gemini Flash suffers from high variability in grading standards across similar questions and students. Analysis reveals wide scoring ranges for equivalent technical implementations of the same concepts, while Claude maintained consistent evaluation patterns. This inconsistency appears to stem from Gemini Flash's limited context memory, which prevents understanding of multi-part questions and their pedagogical interconnections, leading to fragmented evaluation that misses the broader learning objectives.

\subsection{Mock Interview-Based Assessments}

The strong correlation between LLM and human grading, particularly for performant models like Claude-3.7-Sonnet, opens possibilities beyond traditional grading toward novel assessment methodologies. Given that Claude maintains consistent relative student rankings and demonstrates systematic evaluation standards, we explored whether LLMs could facilitate individualized ``oral'' examinations that address several challenges in undergraduate assessment.

Traditional written examinations suffer from several limitations that become acute in large university settings. Students can share solutions, creating fairness concerns and undermining assessment validity. Static examinations provide single-point failure scenarios where students may become stuck on early questions without opportunity for recovery. Additionally, the archaic nature of traditional testing poorly reflects real-world problem-solving scenarios where students would have access to resources and can seek clarification when needed.

Interview-style assessments, while pedagogically superior in many respects \citep{Graesser2001,Huxham2012}, face scalability challenges. In courses with hundreds of students, conducting individual interview-based examinations requires human resources. When distributed across multiple examiner groups, these assessments often suffer from the same inconsistency issues observed in homework grading above, potentially creating unfair evaluation disparities. The flip-classroom approach, while promoted in educational literature \citep{OFlaherty2015}, remains difficult to implement at scale for assessment purposes.

LLM-based interview systems could potentially address these scalability and consistency challenges while preserving the pedagogical benefits of individualized assessment. The system can generate unique question sets from the same course material, eliminating solution sharing concerns.

\subsubsection{System Architecture and Design}

We developed a Socratic-style interview agent based on similar principles as AstroTutor but designed for assessment rather than tutoring. The system was programmed with course materials and instructed to conduct structured ``oral'' examinations covering conceptual understanding, implementation skills, and debugging capabilities. The agent follows a carefully designed protocol that ensures coverage of key learning objectives while adapting to individual student responses and providing appropriate scaffolding when needed.

\textbf{Assessment Framework.} The interview system implements a structured three-category assessment approach. \textit{Conceptual questions} evaluate theoretical understanding by presenting real-world astronomical scenarios and asking students to select and justify appropriate analytical methods. \textit{Implementation tasks} require students to write focused code segments that demonstrate technical proficiency with course algorithms, emphasizing problem-solving approach over syntactic perfection. \textit{Debugging challenges} present students with flawed code implementations, testing their ability to identify logical errors and propose systematic corrections.

\textbf{Scaffolding and Support Mechanisms.} When students encounter difficulties, the system offers graduated support through a hint mechanism that requires explicit student consent. Hints are posed as guiding questions rather than direct answers, maintaining the assessment integrity while preventing complete failure scenarios. This approach mirrors effective human tutoring practices where struggling students receive guidance without compromising the evaluation of their independent capabilities.

The interview format naturally accommodates diverse learning needs and circumstances \citep{Griful-Freixenet2017}. Students with special accommodations can take examinations at their own pace, pause and resume sessions as needed, and access the assessment environment from familiar locations. 

\subsubsection{Pilot Implementation and Evaluation}

For this pilot implementation, we tested the system with a single student who required make-up assessment due to circumstances beyond their control. The assessment was conducted entirely via text-based chat interaction rather than voice, enabling consistent logging of the complete conversation for subsequent evaluation. The student completed sessions covering the major course topics, submitting chat records for evaluation. No time limits were imposed to accommodate the student's circumstances.

The following shows a paraphrased version (for privacy reasons) of a conversation from the interview-based assessment:

\begin{questionbox}[boxcolor1]
\textbf{LLM Examiner:} Looking at your course topics, I'd like you to choose which topic you'd like the first question to be on. Your options are: Random Variables and Bayesian Inference, Summary Statistics, Linear Regression, Classification and Logistic Regression, Principal Component Analysis, or Clustering with Gaussian Mixture Models.\\

\textbf{Student:} I'll choose Random Variables and Bayesian Inference.\\

\textbf{LLM Examiner:} Great choice. Here's a conceptual question for you: Imagine you're analyzing exoplanet transit data. You have a light curve with some noise, and you want to estimate the planet's radius. You can model the transit depth as a function of the planet's radius, but you also have prior knowledge about the typical sizes of planets in similar systems. How would you approach this problem using Bayesian inference, and why is Bayesian inference a suitable choice here?\\

\textbf{Student:} Bayesian modeling differs from subjective interpretations because probability can influence its belief. With a more standard way of thinking, taking into account the `usual' size of a planet's radius in similar systems would be pointless. However, with Bayesian inference, it updates its probability as more data is revealed. If you already have a good idea about the type of planet it might be observing, this could factor into the statistical interpretation of it.\\

\textbf{LLM Examiner:} That's a good start. You've correctly identified that Bayesian inference allows you to incorporate prior knowledge about the planet's radius. To expand on your answer, can you explain how the prior distribution, likelihood function, and posterior distribution would play a role in this specific exoplanet transit scenario?
\end{questionbox}

The session was recorded and subsequently evaluated by both the human instructor and an offline open-weight LLM (Qwen3-235B-A22B) using identical grading criteria. Here, open-weight means the model weights can be downloaded and run locally on our own GPU hardware, rather than accessed through a cloud API. The offline evaluation approach was chosen to protect student privacy. 

The human instructor assessed performance based on demonstrated conceptual understanding, problem-solving approach, technical implementation skills, and ability to articulate complex ideas clearly. Qwen3-235B-A22B received the complete conversation transcript and evaluated using the same rubric, generating detailed feedback across all assessment dimensions. However, the official grade used for course records was based solely on the human instructor's evaluation, with the LLM assessment serving as a comparative validation study.

Both evaluations showed strong consistency in overall performance assessment and specific area identification. The LLM evaluation aligned with the human instructor's judgment regarding the student's conceptual understanding and technical gaps. The interview format revealed learning patterns that traditional written examinations might miss. The real-time feedback mechanism allowed the examiner to probe understanding depths and identify specific misconceptions, providing diagnostic information valuable for both grading and future instruction planning.

This pilot demonstration suggests that LLM-facilitated oral examinations merit further investigation as a potentially scalable solution for individualized assessment in large courses while maintaining the pedagogical benefits of adaptive questioning and real-time feedback. While this single pilot case does not allow generalizable conclusions, subsequent to this study, text-based LLM interview assessments based on this approach have been implemented in a larger introductory astronomical data analysis course at The Ohio State University, where they have been well received by students. The AI quiz systems are publicly available as part of the open course textbook\footnote{\url{https://tingyuansen.github.io/coding_essential_for_astronomers/}} \citep{Ting2025coding}.

\subsubsection{Automated Video Analysis for Academic Integrity}

Screen recording serves as a deterrent mechanism for academic integrity monitoring. We implemented offline LLM-based video analysis with the open-weight Qwen-2.5-VL-72B to detect potential violations in student self-reported Zoom recordings. The analysis identifies suspicious behavior patterns including: (1) extended periods where students appear off-camera, suggesting movement away from the assessment environment, and (2) sustained eye-gaze diversions indicating attention to off-screen resources such as additional browser windows or devices.

The analysis generates timestamped reports identifying intervals warranting closer examination. For example, flagged periods included ``500-510 seconds: Sustained gaze shift---possibly looking off-screen at another browser or document'' and ``850-870 seconds: Sustained gaze shift.''

Following analysis, all flagged segments are extracted and concatenated into a single condensed video for human review, after which original recordings are immediately discarded to protect student privacy. This reduces manual verification time from approximately one hour to one minute of targeted inspection. Human reviewers can efficiently assess whether genuine violations occurred or whether the system detected false positives due to natural movements, lighting conditions, or camera positioning.

While this automated screening approach demonstrates potential scalability for large courses by focusing human review on algorithmically flagged segments rather than requiring comprehensive footage analysis, we deliberately restricted its use to just a single make-up assessment case as described above, using only offline open-weight models run locally on our hardware. This highly limited deployment was intentional, driven by ethical concerns about privacy and surveillance in educational settings \citep{Regan2019,Coghlan2021}. Before considering broader implementation, the academic community needs to engage in thorough discussion about balancing academic integrity verification with student privacy rights and dignity.

\subsection{Ethical Considerations}

The feedback form through which students provided survey responses and homework reflections informed participants that their anonymized responses may be used for educational research purposes. The institutional review board (IRB) was consulted regarding the study, and the development and deployment of AstroTutor were carried out in consultation with the school's IT management. All student data were stored locally and were not shared with third-party services beyond the LLM API calls inherent to the tools and analyses described in this study (i.e., AstroTutor's tutoring functionality and the grading experiments). All data presented in this paper are fully anonymized, with no individually identifiable information disclosed.

While AI as a learning companion represents a less contentious application, AI as a grader and exam setter raises issues that require broader community consensus. We emphasize that all official grading in this course remained entirely under human control, with our LLM analysis serving as a post-hoc comparative study to assess potential capabilities. 

The advantages of LLM-assisted assessment are evident: detailed, personalized feedback can be delivered simultaneously to all students, creating more equitable learning environments. Students receive real-time feedback that identifies knowledge gaps, valuable for formative assessment purposes. The performance stems from LLM's systematic application of evaluation criteria, largely avoiding the potential unconscious bias and inconsistency that research demonstrates affects human graders.

However, human graders retain value through higher-order pedagogical judgment and understanding of individual student circumstances at a more personal level. Beyond technical evaluation capabilities, human accountability — albeit imperfect — remains a more acceptable social norm compared to machine judgment in educational contexts. This parallels ongoing debates around autonomous systems in other domains, where technical capability must be balanced against social acceptance and ethical responsibility.

These broader implications extend beyond the scope of this study. We focus on demonstrating AI as a learning companion while leaving questions of automated grading and assessment implementation to community deliberation. Our findings suggest potential benefits for educational equity and consistency, but widespread implementation requires careful consideration of institutional values, student agency, and pedagogical philosophy \citep{EuropeanCommission2022}.

\section{Recommendations for Implementation}

Based on our findings, we propose several concrete recommendations for educators in astronomy considering similar LLM integration in astronomy education and related STEM disciplines.

\textbf{Require AI Usage Documentation and Reflection.} Require students to document AI usage through reflection questions that encourage critical evaluation rather than simple reporting. Award credit for thoughtful analysis of AI interactions, including failed attempts and limitations encountered. Based on our exploratory observations, this documentation requirement appeared effective in fostering metacognitive awareness and reducing dependency over time.

\textbf{Develop Domain-Specific AI Tutors.} Invest in specialized AI tutors tailored to course content when resources permit. While more expensive than general-purpose models, domain-specific tools like AstroTutor provide more reliable information and better pedagogical alignment with learning objectives. Students valued the specialized knowledge while still utilizing commercial models for complementary purposes.

\textbf{Guide Multi-Tool Usage Rather Than Restrict Access.} Recognize that students will use multiple AI tools for different purposes. Rather than restricting access, guide students toward understanding when different tools are most appropriate and effective. Introduce surveys of available models at the beginning of courses to help students make informed tool choices.

\textbf{Teach Modern Development Environments and API Usage.} Provide more extensive introductions to Integrated Development Environments (IDEs)—software applications that combine text editors, debugging tools, and project management features in a single interface for writing and testing code—particularly for modern AI-integrated development environments like Cursor AI or GitHub Copilot, to better prepare students for professional practice. Teaching students to use Application Programming Interfaces (APIs)—direct computer-to-computer communication methods that allow automated interactions with AI systems—eliminates limitations stemming from students only knowing Graphical User Interfaces (GUIs), the familiar point-and-click web interfaces like ChatGPT's website. API access enables more sophisticated AI interactions through automated workflows, batch processing, and integration with coding environments.

\textbf{Consider LLM-Assisted Assessment Methods.} Consider implementing LLM-assisted grading for consistent, detailed feedback while maintaining human oversight for official evaluation. Explore interview-based assessments facilitated by AI for personalized evaluation at scale, particularly for make-up examinations or alternative assessment needs.

Several improvements could enhance future iterations. Expanding agentic AI tutors with more extensive astronomy-focused code bases would improve domain-specific assistance quality. Introducing AI literacy concepts early, including prompt engineering and output verification strategies, would accelerate student proficiency development. Creating structured opportunities for peer sharing of effective AI strategies could enhance community learning around these tools.

\section{Conclusion}

This study represents an examination of LLM integration in astronomy education through an advanced undergraduate course in astro-statistics targeting final-year students preparing for graduate school in astronomy. The course emphasized problem-solving through machine learning and statistical techniques applied to real astronomical research problems, providing an ideal context for exploring AI integration in computationally intensive STEM education. While conducted with a small class size, the outcomes provide insight to warrant broader consideration.

Our exploratory study yielded several findings that are suggestive, though not causally conclusive, regarding common assumptions about AI in education. The development and deployment of AstroTutor suggested that domain-specific AI tutors can provide reliable, contextually appropriate assistance while reducing hallucinations common in general-purpose models. Students valued this specialized tool as a complement to commercial LLMs, developing strategies for using different tools for different purposes.

The documentation requirement and encouragement of critical thinking appeared effective in fostering reflective AI usage. Rather than creating dependency, structured LLM integration was associated with decreased reliance over time as students reported developing independent problem-solving capabilities. This trajectory suggests that LLMs can serve as effective learning scaffolds when properly implemented with appropriate reflection mechanisms.

Students demonstrated evolution in their approach to AI tools, progressing from basic assistance-seeking to strategic, critical engagement. Students approached LLMs as learning tools rather than shortcuts, attempting problems independently first and seeking targeted assistance for specific obstacles. This observed approach provides preliminary evidence against common concerns about AI undermining educational objectives, though we note the limitations of self-reported data in drawing definitive conclusions.

Students developed AI literacy skills through their LLM interactions, learning to evaluate different tools' strengths and limitations, formulate effective prompts, verify outputs against authoritative sources, choose appropriate tools for specific tasks, and balance efficiency with learning goals. These AI proficiency skills will likely remain valuable throughout their careers as AI continues to evolve in scientific and technical fields.

Our comparative grading analysis revealed strong correlation between LLM and human evaluation. The systematic consistency of LLM grading offers potential solutions to scalability and fairness challenges in large courses, while the pilot interview-based assessment provided a preliminary demonstration of novel possibilities for individualized evaluation. However, ethical considerations require community consensus before widespread implementation of automated assessment.

Several limitations should be noted when interpreting these results. The study involved 12 self-selected final-year students in a specialized astrostatistics course at a single institution, with post-course survey data available from only 7 respondents. We note that this class size is typical for upper-level astronomy courses---even at Ohio State, which has one of the five largest undergraduate astronomy programs in the United States, this is the expected enrollment for a specialized final-year course. AI tutoring tools like AstroTutor are particularly well suited to such advanced courses, where students face difficult, research-oriented material but class sizes are inherently small. Nonetheless, this small, highly motivated cohort is not representative of broader student populations, and the findings may not transfer directly to introductory astronomy courses, larger class sizes, or less computationally oriented curricula. The absence of a control group or pre/post assessment design means that observed outcomes cannot be causally attributed to the LLM intervention. 

However, the reliance on self-reporting was a deliberate pedagogical and ethical choice: the documentation requirement was itself an integral part of the learning intervention, designed to foster metacognitive awareness of LLM limitations, and withholding LLM access from a control group would have been difficult to justify ethically in a course where LLM literacy was a stated learning objective. Nevertheless, to our best knowledge, this represents the first dedicated report on structured LLM integration in astronomy education, and we believe the detailed implementation description, student interaction patterns, and practical guidelines provide a valuable starting point for the astronomy education community, motivating future controlled studies with larger and more diverse samples.

This case study suggests that thoughtfully integrated LLMs can enhance astronomy education when students are encouraged to use them as learning tools rather than answer providers, and when instructors create structures that promote reflection on effective AI integration strategies. The key finding is that structured encouragement of AI use, combined with transparency requirements and domain-specific tools, can enhance astronomy education without compromising learning objectives, preparing students for an AI-integrated future while maintaining the rigor and independence that characterize effective STEM education.

\appendix
\section{System Configuration and Reproducibility}
\label{appendix:config}

The full source code for AstroTutor, including all configuration files, prompt templates, and embedding parameters, is publicly available at \url{https://github.com/tingyuansen/computeastro-tutor}. The repository contains the RAG agent system prompts, moderator agent configuration, retrieval parameters (e.g., top-$k$ settings, similarity thresholds), and ChromaDB vector store setup. The system was built using Google's Gemini-2.0-Flash (later upgraded to Gemini-2.5-Flash during the semester) with text-embedding-004 for vector embeddings. The LLM grading code, which uses Claude-3.7-Sonnet and Gemini-2.5-Flash, is also included in the repository. We refer readers to the repository documentation for complete implementation details.

\bibliographystyle{aasjournal}
\bibliography{manuscript.bib}

\end{document}